\documentclass[onecolumn,showpacs,preprintnumbers,amsmath,amssymb,eqsecnum]{revtex4}
\usepackage{dcolumn}
\usepackage{bm}
\usepackage[dvips]{graphicx}
\usepackage{wrapfig}
\usepackage{psfrag}
\usepackage{color}
\begin{document}


devoted to memory of Alexei Alexandrovich Starobinsky

\def\sh{\mathop{\rm sh}\nolimits}
\def\ch{\mathop{\rm ch}\nolimits}
\def\var{\mathop{\rm var}}\def\exp{\mathop{\rm exp}\nolimits}
\def\Re{\mathop{\rm Re}\nolimits}
\def\Sp{\mathop{\rm Sp}\nolimits}
\def\kp{\mathop{\text{\ae}}\nolimits}
\def\bk{{\bf {k}}}
\def\bp{{\bf {p}}}
\def\bq{{\bf {q}}}
\def\lra{\mathop{\longrightarrow}}
\def\Const{\mathop{\rm Const}\nolimits}
\def\sh{\mathop{\rm sh}\nolimits}
\def\ch{\mathop{\rm ch}\nolimits}
\def\var{\mathop{\rm var}}
\def\mK{\mathop{{\mathfrak {K}}}\nolimits}
\def\mR{\mathop{{\mathfrak {R}}}\nolimits}
\def\mv{\mathop{{\mathfrak {v}}}\nolimits}
\def\mV{\mathop{{\mathfrak {V}}}\nolimits}
\def\mD{\mathop{{\mathfrak {D}}}\nolimits}
\def\mN{\mathop{{\mathfrak {N}}}\nolimits}
\def\mS{\mathop{{\mathfrak {S}}}\nolimits}
\def\mP{\mathop{{\mathfrak {P}}}\nolimits}
\def\mL{\mathop{{\mathfrak {L}}}\nolimits}
\def\ml{\mathop{{\mathfrak {l}}}\nolimits}

\newcommand\ve[1]{{\mathbf{#1}}}

\def\Re{\mbox {Re}}
\newcommand{\Z}{\mathbb{Z}}
\newcommand{\R}{\mathbb{R}}
\def\mK{\mathop{{\mathfrak {K}}}\nolimits}
\def\mk{\mathop{{\mathfrak {k}}}\nolimits}
\def\mR{\mathop{{\mathfrak {R}}}\nolimits}
\def\mv{\mathop{{\mathfrak {v}}}\nolimits}
\def\mV{\mathop{{\mathfrak {V}}}\nolimits}
\def\mD{\mathop{{\mathfrak {D}}}\nolimits}
\def\mN{\mathop{{\mathfrak {N}}}\nolimits}
\def\ml{\mathop{{\mathfrak {l}}}\nolimits}
\def\mf{\mathop{{\mathfrak {f}}}\nolimits}
\newcommand{\ccm}{{\cal M}}
\newcommand{\cE}{{\cal E}}
\newcommand{\cV}{{\cal V}}
\newcommand{\cI}{{\cal I}}
\newcommand{\cR}{{\cal R}}
\newcommand{\cK}{{\cal K}}
\newcommand{\cH}{{\cal H}}
\newcommand{\cW}{{\cal W}}
\newcommand{\cL}{{\cal L}}

\def\br{\mathop{{\bf {r}}}\nolimits}
\def\bS{\mathop{{\bf {S}}}\nolimits}
\def\bA{\mathop{{\bf {A}}}\nolimits}
\def\bJ{\mathop{{\bf {J}}}\nolimits}
\def\bn{\mathop{{\bf {n}}}\nolimits}
\def\bg{\mathop{{\bf {g}}}\nolimits}
\def\bv{\mathop{{\bf {v}}}\nolimits}
\def\be{\mathop{{\bf {e}}}\nolimits}
\def\bp{\mathop{{\bf {p}}}\nolimits}
\def\bz{\mathop{{\bf {z}}}\nolimits}
\def\bbf{\mathop{{\bf {f}}}\nolimits}
\def\bb{\mathop{{\bf {b}}}\nolimits}
\def\ba{\mathop{{\bf {a}}}\nolimits}
\def\bx{\mathop{{\bf {x}}}\nolimits}
\def\by{\mathop{{\bf {y}}}\nolimits}
\def\br{\mathop{{\bf {r}}}\nolimits}
\def\bs{\mathop{{\bf {s}}}\nolimits}
\def\bH{\mathop{{\bf {H}}}\nolimits}
\def\bk{\mathop{{\bf {k}}}\nolimits}
\def\be{\mathop{{\bf {e}}}\nolimits}
\def\bnul{\mathop{{\bf {0}}}\nolimits}
\def\bq{{\bf {q}}}

\newcommand{\oV}{\overline{V}}
\newcommand{\vkp}{\varkappa}
\newcommand{\os}{\overline{s}}
\newcommand{\opsi}{\overline{\psi}}
\newcommand{\ov}{\overline{v}}
\newcommand{\oW}{\overline{W}}
\newcommand{\oPhi}{\overline{\Phi}}

\def\mI{\mathop{{\mathfrak {I}}}\nolimits}
\def\mA{\mathop{{\mathfrak {A}}}\nolimits}

\def\st{\mathop{\rm st}\nolimits}
\def\tr{\mathop{\rm tr}\nolimits}
\def\sign{\mathop{\rm sign}\nolimits}
\def\d{\mathop{\rm d}\nolimits}
\def\const{\mathop{\rm const}\nolimits}
\def\diag{\mathop{\rm diag}\nolimits}
\def\O{\mathop{\rm O}\nolimits}
\def\Spin{\mathop{\rm Spin}\nolimits}
\def\exp{\mathop{\rm exp}\nolimits}
\def\SU{\mathop{\rm SU}\nolimits}
\def\mU{\mathop{{\mathfrak {U}}}\nolimits}
\newcommand{\cU}{{\cal U}}
\newcommand{\cD}{{\cal D}}

\def\mI{\mathop{{\mathfrak {I}}}\nolimits}
\def\mA{\mathop{{\mathfrak {A}}}\nolimits}
\def\mU{\mathop{{\mathfrak {U}}}\nolimits}

\def\st{\mathop{\rm st}\nolimits}
\def\tr{\mathop{\rm tr}\nolimits}
\def\sign{\mathop{\rm sign}\nolimits}
\def\d{\mathop{\rm d}\nolimits}
\def\const{\mathop{\rm const}\nolimits}
\def\O{\mathop{\rm O}\nolimits}
\def\Spin{\mathop{\rm Spin}\nolimits}
\def\exp{\mathop{\rm exp}\nolimits}

\title{An alternative idea about the source of baryon asymmetry in the Universe}

\author {S.N. Vergeles\vspace*{4mm}\footnote{{e-mail:vergeles@itp.ac.ru}}}

\affiliation{Landau Institute for Theoretical Physics,
Russian Academy of Sciences,
Chernogolovka, Moscow region, 142432 Russia \linebreak
and   \linebreak
Moscow Institute of Physics and Technology, Department
of Theoretical Physics, Dolgoprudnyj, Moskow region,
141707 Russia}

\begin{abstract}
The paper proposes an alternative scenario for the emergence of baryon asymmetry in the Universe. This scenario is realized in the lattice gravity model associated with the Dirac field as follows. At ultra-high temperatures of the Grand Unification order $T_c\sim10^{18}$ GeV and higher, the system is in a PT-symmetric phase. But when the temperature decreases, a phase transition to an asymmetric phase occurs, in which a non-zero tetrad appears, that is, space-time with the Minkowski metric, and the system's wave function splits into two: $|\rangle=|+\rangle+|-\rangle$. The fields of tetrads in states $|+\rangle$ and $|-\rangle$ differ in sign. At the very first moment of time with a duration of the order of the Planck time, a transition of fermions between these states is possible. These transitions in different parts of space are not correlated with each other. Therefore, the final asymmetry of the fermion charge between these states is relatively extremely small and it is preserved in time, since the interaction of the states $|+\rangle$ and $|-\rangle$ ceases at times greater than the Planck time.
\end{abstract}


\maketitle

\section{Introduction}

The problem of the existence of baryon asymmetry in the Universe has not been solved at present.
Much information, as well as a significant number of references on this issue, are contained in the reviews \cite{rubakov1996electroweak,vergeles2023general}.
Here we briefly note only the following facts.

1) It is usually assumed that an electro-weak phase transition that spontaneously breaks the $SU(2)_L\times U(1)_Y$ symmetry is a first-order transition. The accumulation of baryons occurs in close proximity to the point of this transition in the asymmetric phase, and embrions and bubbles of the asymmetric phase accelerate the process \cite{rubakov1996electroweak,kuzmin1985anomalous}.

2) All approaches to solving the problem (for example, the use of the Kobayashi–Maskawa mechanism in the Standard Model or the so-called  sphaleron baryogenesis  proposed by Kuzmin,
Rubakov, and Shaposhnikov \cite{kuzmin1985anomalous}) lead to estimates of the average baryon density that is many orders of magnitude lower than observed ($<10^{-5}$).

Here we present an alternative scenario for the accumulation of baryon asymmetry at the earliest stage of the development of the Universe. This scenario is possible in the lattice theory of gravity. The author's works \cite{vergeles2006one,vergeles2015wilson,vergeles2017note,vergeles2017fermion,
vergeles2021note,vergeles2021domain,vergeles2023another} studied the lattice theory of gravity associated with Dirac fields. In particular, in work \cite{vergeles2021note}
a discrete $\Z_2$-symmetry (called $PT$-symmetry) was described, mutually interchanging the Dirac and its conjugate fields. In work \cite{vergeles2023another}
it was proven that at ultrahigh temperatures this $PT$ symmetry is not broken, but when the temperature decreases in the inflation phase, the $PT$ symmetry is broken. This work shows that in the asymmetric phase in close proximity to the phase transition point in temperature and time, the symmetric wave function (w.f.) of the Universe breaks up into a superposition of two w.fs. $|+\rangle$ and $|-\rangle$. During this small Planck time $\tau\sim t_P\approx\sqrt{8\pi G\hbar/c^5}\approx10^{-43}$ sec. matrix element
$\langle+|{\cal H}_{\Psi}|-\rangle\neq0$.
Since the Hamiltonian preserves the total number of fermions, the indicated inequality contains the amplitudes of fermion transitions from $|-\rangle$ to $|+\rangle$, and vice versa. For times $t>\tau$ we have $\langle+|\hat{\cal O}|-\rangle=0$ for all local operators $\hat{\cal O}$. Therefore, the imbalance of fermions between the states $|+\rangle$ and $|-\rangle$ accumulated over time $\tau$ is preserved, which is the source of baryon asymmetry.

Let us try to explain the physics of the phenomenon using an extremely simple example from one-dimensional quantum mechanics. Let the single-particle Hamiltonian be ${\cal H}=-(1/2m)\d^2/\d x^2-\varkappa[\delta(x-a)+\delta(x+a)]$. This Hamiltonian is symmetric under the parity transformation $x\longrightarrow-x$, but of the two bound states, one $\phi_s(x)$ is even, and the other $\phi_a(x)$ is odd. Their linear combinations $\phi_{\pm}(x)=\phi_s(x,t)\pm\phi_a(x,t)$ are not stationary, transform into each other under parity transformation and describe the wave functions of states in which the particle is near $x=\pm a$, respectively. If a particle is in the state $\phi_+(x)$, then after some time it will be in the state $\phi_-(x)$, and so on. If at some point in time an impenetrable barrier appears at the point $x=0$, then the probabilities of detecting a particle in the region $x>0$ and $x<0$ $"\mbox{freeze}"$, not being, generally speaking, equal. Something similar happens in the model under study. Although the action of the system is $PT$-invariant, the states $|+\rangle$ and $|-\rangle$ are not, and they transform into each other under $PT$-transformation. The $\phi_{\pm}$ states are simplified analogues of the $|\pm\rangle$ states. The latter are a superposition of many states, including those with different values of Fermi particles. The above-mentioned $"\mbox{impenetrable barrier}"$ between the states $\phi_+$ and $\phi_-$ in the model under consideration arises spontaneously and within a minimum time $t_P$.

The consideration is of a model nature.

It should be noted that the present work is not pioneering in the sense that gravity as a source of baryon asymmetry was first considered in \cite{davoudiasl2004gravitational}. In this work, an additional term $\{(1/M^2_*)\sqrt{-g}(\partial_{\mu}\mR)J^{\mu}_B\}$ was introduced into the Lagrangian, violating $CP$-invariance. Here $\mR$ is the scalar curvature, $J^{\mu}_B$ is the baryon non-conserved current and $M^2_*$ is the cutoff parameter of the effective theory. Among the latest works developing this idea, we note the works \cite{arbuzova2023gravitational,mishra2023constraining}, in which one can find many references on this topic. Without going into details, we point out the fundamental difference between our approach to the problem and the approach in \cite{davoudiasl2004gravitational,arbuzova2023gravitational,mishra2023constraining}: in the latter approach, the continuum theory of gravity is studied, into which the $CP$ non-invariant term is initially introduced; in our approach, the lattice, i.e. regularized, theory of gravity is studied, which is invariant with respect to all symmetries in the high-temperature phase, but as a result of the spontaneous phase transition to the low-temperature phase with the loss of $PT$ symmetry, baryon asymmetry is generated (see the text below).

To make the article easier to read, the next Section provides a definition of the lattice gravity model that is being studied here.

\section{Definition of lattice theory of gravity}

Consider the orientable abstract 4-dimensional simplicial complex  $\mK$. We recommend the book \cite{pontryagin1976basics}, \S\S 2,4 for an introduction to the definition of abstract simplicial complexes. Suppose that any of its 4-simplexes belongs to such a finite (or infinite) sub-complex ${\mK}'\in\mK$  which has a geometric realization in  $\R^4$ topologically equivalent to a disk  without cavities. The vertices are designated as
$a_{\cV}$, the indices ${\cV}=1,2,\dots,\,{\mN}^{(0)}\rightarrow\infty$ and ${\cW}$ enumerate the vertices and 4-simplices, correspondingly. It is necessary to use
the local enumeration of the vertices $a_{\cV}$ attached to a given
4-simplex: the all five vertices of a 4-simplex with index ${\cW}$
are enumerated as $a_{{\cV}_{({\cW})i}}$, $i=1,2,3,4,5$. The later notations with extra low index  $({\cW})$
indicate that the corresponding quantities belong to the
4-simplex with index ${\cW}$. Of course, these same quantities  also belong to another 4-simplex with index  ${\cW}'$, and 4-simplexes with indices  ${\cW}$ and  ${\cW}'$ must be adjacent. Let us denote
$\varepsilon_{{\cV}_{({\cW})1}{\cV}_{({\cW})2}{\cV}_{({\cW})3}{\cV}_{({\cW})4}{\cV}_{({\cW})5}}=\pm 1$
the Levi-Civita symbol. The upper (lower) sign depends on the orientation of the 4-simplex
$s^4_{\cW}=a_{{\cV}_{({\cW})1}}a_{{\cV}_{({\cW})2}}
a_{{\cV}_{({\cW})3}}a_{{\cV}_{({\cW})4}}a_{{\cV}_{({\cW})5}}$.
An element of the compact group $\Spin(4)$ and an element of the Clifford algebra
\begin{gather}
\Omega_{{\cV}_1{\cV}_2}=\Omega^{-1}_{{\cV}_2{\cV}_1}=\exp\left(
\omega_{{\cV}_1{\cV}_2}\right)=
\exp\left(\frac{1}{2}\sigma^{ab}
\omega^{ab}_{{\cV}_1{\cV}_2}\right)\in\Spin(4),
\nonumber \\
\sigma^{ab}\equiv\frac{1}{4}[\gamma^a,\gamma^b], \quad \gamma^a\gamma^b+\gamma^b\gamma^a=2\delta^{ab}, \quad  a=1,2,3,4, \quad
\gamma^5\equiv\gamma^1\gamma^2\gamma^3\gamma^4=(\gamma^5)^{\dag},
\nonumber \\
\hat{e}_{{\cV}_1{\cV}_2}\equiv e^a_{{\cV}_1{\cV}_2}\gamma^a\equiv
-\Omega_{{\cV}_1{\cV}_2}\hat{e}_{{\cV}_2{\cV}_1}\Omega_{{\cV}_1{\cV}_2}^{-1},
\nonumber \\
|e_{{\cV}_1{\cV}_2}|<1, \quad |e_{{\cV}_1{\cV}_2}|\equiv\sqrt{\sum_a(e^a_{{\cV}_1{\cV}_2})^2}
\label{Variables_Grav}
\end{gather}
are assigned for each oriented 1-simplex $a_{{\cV}_1}a_{{\cV}_2}$.
The conjecture is that the set of variables $\{\Omega,\,\hat{e}\}$  is an independent set of dynamic variables.
Fermionic degrees of freedom (Dirac spinors) are assigned to each vertex of the complex:
\begin{gather}
\Psi^{\dag}_{\cV}, \quad \Psi_{\cV}.
\label{Variables_Ferm}
\end{gather}
The set of variables $\{\Psi^{\dag},\,\Psi\}$ is a set of mutually independent variables, and the spinors $\Psi^{\dag}_{\cV}$ and $\Psi_{\cV}$ are in mutual involution (or anti-involution) relative to Hermitian conjugation operation.

Consider a model with an action
\begin{gather}
\mA=\mA_g+\mA_{\Psi}+\mA_{\Lambda_0}.
\label{Action_4D}
\end{gather}
Here $\mA_g$ and $\mA_{\Psi}$ are the actions of pure gravity and Dirac field, correspondingly:
\begin{gather}
\mA_g=-\frac{1}{5!\cdot2\cdot l_P^2}\sum_{\cW}\sum_{\sigma}
\varepsilon_{\sigma({\cV}_{({\cW})1})\sigma({\cV}_{({\cW})2})
\sigma({\cV}_{({\cW})3})\sigma({\cV}_{({\cW})4})\sigma({\cV}_{({\cW})5})}
\nonumber \\
\times\tr\gamma^5\bigg\{
\Omega_{\sigma({\cV}_{({\cW})5})\sigma({\cV}_{({\cW})1})}
\Omega_{\sigma({\cV}_{({\cW})1})\sigma({\cV}_{({\cW})2})}\Omega_{\sigma({\cV}_{({\cW})2})\sigma({\cV}_{({\cW})5})}
\hat{e}_{\sigma({\cV}_{({\cW})5})\sigma({\cV}_{({\cW})3})}
\hat{e}_{\sigma({\cV}_{({\cW})5})\sigma({\cV}_{({\cW})4})}\bigg\}.
\label{Latt_Action_Grav}
\end{gather}
Each $\sigma$ is one of 5! vertex permutations ${\cV}_{({\cW})i}\longrightarrow\sigma(
{\cV}_{({\cW})i})$.
\begin{gather}
\mA_{\Psi}=\frac{1}{5\cdot24^2}\sum_{\cW}\sum_{\sigma}
\varepsilon_{\sigma({\cV}_{({\cW})1})\sigma({\cV}_{({\cW})2})
\sigma({\cV}_{({\cW})3})\sigma({\cV}_{({\cW})4})\sigma({\cV}_{({\cW})5})}
\nonumber \\
\times\tr\gamma^5\bigg\{ \hat{\Theta}_{\sigma({\cV}_{({\cW})5})\sigma({\cV}_{({\cW})1})}
\hat{e}_{\sigma({\cV}_{({\cW})5})\sigma({\cV}_{({\cW})2})}
\hat{e}_{\sigma({\cV}_{({\cW})5})\sigma({\cV}_{({\cW})3})}
\hat{e}_{\sigma({\cV}_{({\cW})5})\sigma({\cV}_{({\cW})4})}\bigg\},
\label{Latt_Action_Ferm}
\end{gather}
\begin{gather}
\hat{\Theta}_{{\cV}_1{\cV}_2}
\equiv\Theta^a_{{\cV}_1{\cV}_2}\gamma^a=\hat{\Theta}_{{\cV}_1{\cV}_2}^{\dag},  \quad
\Theta^a_{\cV_1\cV_2}=\frac{i}{2}\left(\Psi^{\dag}_{\cV_1}\gamma^a\Omega_{{\cV}_1{\cV}_2}\Psi_{\cV_2}-
\Psi^{\dag}_{\cV_2}\Omega_{{\cV}_2{\cV}_1}\gamma^a\Psi_{\cV_1}\right).
\label{Dirac_Form}
\end{gather}
 It is easy to check that (compare with (\ref{Variables_Grav}))
\begin{gather}
\hat{\Theta}_{{\cV}_1{\cV}_2}
\equiv-\Omega_{{\cV}_1{\cV}_2}\hat{\Theta}_{{\cV}_2{\cV}_1}
\Omega_{{\cV}_1{\cV}_2}^{-1}.
\label{Dir_Bil_Form_Trans}
\end{gather}
The contribution to the lattice action from the cosmological constant has the form
\begin{gather}
\mA_{\Lambda_0}=-\frac{1}{5!\cdot12}\cdot\frac{\Lambda_0}{l_P^2}\varepsilon_{abcd}\sum_{\cW}\sum_{\sigma}
\varepsilon_{\sigma({\cV}_{({\cW})1})\sigma({\cV}_{({\cW})2})
\sigma({\cV}_{({\cW})3})\sigma({\cV}_{({\cW})4})\sigma({\cV}_{({\cW})5})}
\nonumber \\
\times e^a_{\sigma({\cV}_{({\cW})5})\sigma({\cV}_{({\cW})1})}
e^b_{\sigma({\cV}_{({\cW})5})\sigma({\cV}_{({\cW})2})}
e^c_{\sigma({\cV}_{({\cW})5})\sigma({\cV}_{({\cW})3})}
e^d_{\sigma({\cV}_{({\cW})5})\sigma({\cV}_{({\cW})4})}.
\label{Latt_Action_Lambda}
\end{gather}
The partition function is represented by integral
\begin{gather}
Z=\prod_{1-\mbox{simplices}}\int_{|e_{{\cV}_1{\cV}_2}|<1}\prod_a\d e^a_{{\cV}_1{\cV}_2}
\int\d\mu\{\Omega_{{\cV}_1{\cV}_2}\}
\prod_{\cV}\int\d\Psi^{\dag}_{\cV}\d\Psi_{\cV}\exp(\mA).
\label{Partition_function}
\end{gather}
The action (\ref{Action_4D}) is invariant relative to the gauge transformations
\begin{gather}
\tilde{\Omega}_{{\cV}_1{\cV}_2}
=S_{{\cV}_1}\Omega_{{\cV}_1{\cV}_2}S^{-1}_{{\cV}_2}, \quad
\tilde{\hat{e}}_{{\cV}_1{\cV}_2}=S_{{\cV}_1}\,\hat{e}_{{\cV}_1{\cV}_2}\,S^{-1}_{{\cV}_1}, \quad
\tilde{\Psi}_{\cV}=S_{\cV}\Psi_{\cV}, \quad \tilde{\Psi^{\dag}}_{\cV}=\Psi_{\cV}^{\dag}S_{\cV}^{-1}, \quad  S_{\cV}\in\Spin(4).
\label{Gauge_Trans}
\end{gather}
Verification of this fact is facilitated by using the relation (compare with the relation for
$\hat{e}_{{\cV}_1{\cV}_2}$ in (\ref{Gauge_Trans}))
\begin{gather}
\tilde{\hat{\Theta}}_{{\cV}_1{\cV}_2}=S_{{\cV}_1}\hat{\Theta}_{{\cV}_1{\cV}_2}S^{-1}_{{\cV}_1},
\label{Teta_Gauge_Trans}
\end{gather}
which follows directly from (\ref{Gauge_Trans}).

The considered lattice model is invariant with respect to the global discrete $\Z_2$-symmetry, which is an analog of the combined PT-symmetry. Let $\hat{\cal U}_{PT}$ denote the operator of this transformation. Then the transformed dynamic variables are expressed in terms of the original variables as follows:
\begin{gather}
\hat{\cal U}_{PT}^{-1}\Psi_{\cV}\hat{\cal U}_{PT}=U_{PT}\left(\Psi^{\dag}_{\cV}\right)^t,
\quad
\hat{\cal U}_{PT}^{-1}\Psi^{{\dag}}_{\cV}\hat{\cal U}_{PT}=-\left(\Psi_{\cV}\right)^tU^{-1}_{PT}, \quad U_{PT}=\gamma^1\gamma^3
\nonumber \\
\hat{\cal U}_{PT}^{-1}e^a_{{\cV}_1{\cV}_2}\hat{\cal U}_{PT}=-e^{a}_{{\cV}_1{\cV}_2}, \quad
\hat{\cal U}_{PT}^{-1}\omega^{ab}_{{\cV}_1{\cV}_2}\hat{\cal U}_{PT}=\omega^{ab}_{{\cV}_1{\cV}_2}.
\label{PT_transform}
\end{gather}
Here the superscript $"t"$ denotes the matrix transposition of the Dirac matrices and spinors.
We have:
\begin{gather}
U^{-1}_{PT}\gamma^aU_{PT}=(\gamma^a)^t, \quad
U^{-1}_{PT}\sigma^{ab}U_{PT}=-(\sigma^{ab})^t.
\label{PT_trans_Dir_Alg}
\end{gather}
It follows from (\ref{PT_transform}) and (\ref{PT_trans_Dir_Alg}) that
\begin{gather}
 U^{-1}_{PT}\Omega_{{\cV}_1{\cV}_2}U_{PT}=\left(\Omega_{{\cV}_2{\cV}_1}\right)^t,
\label{PT_trans_Conn}
\end{gather}
as well as
\begin{gather}
\hat{\cal U}_{PT}^{-1}\Theta^a_{{\cV}_1{\cV}_2}\hat{\cal U}_{PT}=-\Theta^a_{{\cV}_1{\cV}_2}.
\label{PT_trans_Ferm_Bil_Fjrm}
\end{gather}

Now let us pass on to the limit of slowly varying fields, that is, to the limit of slowly changing fields when moving along the lattice. In this limit, the action (\ref{Action_4D}) transforms into the well-known continuous action of gravity in the form of the Palatini and Dirac fields minimally coupled to gravity, plus a contribution from the cosmological constant. This transition have meaning together with the transition to Minkowski signature. As a result the compact gauge group $\Spin(4)$ transforms into the non-compact group $\Spin(3,1)$.

In the rest of this section, all lattice variables in the case of the Euclidean signature are primed. For field variables in the case of the Minkowski signature, the old notation is used.

Firstly let us perform the following deformations of integration contours  in integral (\ref{Partition_function}):
\begin{gather}
{\omega'}_{{\cV}_1{\cV}_2}^{4\alpha}= i\omega^{0\alpha}_{{\cV}_1{\cV}_2}, \quad
{\omega'}_{{\cV}_1{\cV}_2}^{\alpha\beta}=-\omega_{{\cV}_1{\cV}_2}^{\alpha\beta},
\nonumber \\
{e'}^4_{{\cV}_1{\cV}_2}= e^0_{{\cV}_1{\cV}_2}, \quad {e'}^{\alpha}_{{\cV}_1{\cV}_2}= ie^{\alpha}_{{\cV}_1{\cV}_2}.
\label{Variables_Trans_Mink}
\end{gather}
The variables $\omega^{ab}_{{\cW}ij}$, $e^a_{{\cW}ij}$ are real quantities for Minkowski signature,
and the indices take on the values
\begin{gather}
a,\,b\ldots=0,1,2,3, \quad \alpha,\,\beta,\ldots=1,2,3.
\label{Mink_Ind}
\end{gather}
The Dirac matrices are transformed as follows:
\begin{gather}
{\gamma'}^4=\gamma^0, \quad {\gamma'}^{\alpha}= i\gamma^{\alpha}, \quad
{\gamma'}^5=\gamma^5=i\gamma^0\gamma^1\gamma^2\gamma^3,
\nonumber \\
\frac12(\gamma^a\gamma^b+\gamma^b\gamma^a)=\eta^{ab}=\diag(1,\,-1,\,-1,\,-1), \quad
\tr\gamma^5\gamma_a\gamma_b\gamma_c\gamma_d=4i\varepsilon_{abcd}, \quad \varepsilon_{0123}=1.
\label{Mink_Dirac_matr}
\end{gather}
Thus, for $\sigma^{ab}=(1/4)[\gamma^a,\,\gamma^b]$ we get
\begin{gather}
{\sigma'}^{4\alpha}= i\sigma^{0\alpha}, \quad {\sigma'}^{\alpha\beta}=-\sigma^{\alpha\beta}.
\label{Mink_Spin_Matr}
\end{gather}
Raising and lowering indices $a,b,\ldots$ is done using tensors $\eta^{ab}$ and $\eta_{ab}$, respectively.
 As a result of (\ref{Variables_Trans_Mink})-(\ref{Mink_Spin_Matr}) we have
\begin{gather}
{\omega'}_{{\cV}_1{\cV}_2}=\frac12\omega^{ab}_{{\cV}_1{\cV}_2}\,\sigma_{ab}
\equiv\omega_{{\cV}_1{\cV}_2},
\quad
{\hat{e}'}_{{\cV}_1{\cV}_2}=\gamma_ae^a_{{\cV}_1{\cV}_2}\equiv\hat{e}_{{\cV}_1{\cV}_2},
\label{Mink_5}
\end{gather}
and also
\begin{gather}
{\Omega'}_{{\cV}_1{\cV}_2}=\exp\left(\frac12{\omega'}_{{\cV}_1{\cV}_2}^{ab}{\sigma'}^{ab}\right)=
\exp\left(\frac12\omega_{{\cV}_1{\cV}_2}^{ab}
\sigma_{ab}\right)\equiv\Omega_{{\cV}_1{\cV}_2}\in\Spin(3,1).
\label{Mink_6}
\end{gather}
We see that the holonomy elements $\Omega_{{\cV}_1{\cV}_2}$ become the elements of the non-compact group $\Spin(3,1)$.

Dirac variables are transformed according to
\begin{gather}
\Psi'_{\cV}=\Psi_{\cV}, \quad  {\Psi'}^{\dag}_{\cV}=\Psi_{\cV}^{\dag}\gamma^0=\overline{\Psi}_{\cV}.
\label{Mink_Dir}
\end{gather}
in passing to the Minkowski signature.

The transition to the long-wave limit is possible for such field configurations that change quite slowly during transitions from simplex to simplex, that is, during small or significant movements along the lattice. This rule applies to any lattices. In our theory, it is precisely at the stage of transition to the long-wave limit that the need arises to introduce local coordinates. Local coordinates are the markers of the lattice vertices. Consider some 4D subcomplex ${\mK}'\in\mK$ with the trivial topology of a four-dimensional disk and a geometric realization in $\R^4$. Thus, each vertex of the subcomplex acquires coordinates $x^{\mu}$, which are the coordinates of the vertex's image in $\R^4$:
\begin{gather}
x^{\mu}_{\cV}\equiv x^{\mu}(a_{\cV}),
 \qquad \ \mu=1,\,2,\,3,\,4.
\label{intr110}
\end{gather}
At this stage the coordinates are dimensionless.
Consider a certain simplex  $s^4_{\cW}\in{\mK}'$. We denote all five vertices of this 4-simplex as ${\cV}_i$, $i=1,2,3,4$ and ${\cV}_m\neq {\cV}_i$. The properties of the geometric realization are such that four infinitely small vectors
\begin{gather}
\d x^{\mu}_{{\cV}_m{\cV}_i}\equiv x^{\mu}_{{\cV}_i}-x^{\mu}_{{\cV}_m}=
-\d x^{\mu}_{{\cV}_i{\cV}_m}\in\R^4 ,  \quad
i=1,\,2,\,3,\,4
\label{intr120}
\end{gather}
are linearly independent. The differentials of coordinates
(\ref{intr120}) correspond to one-dimensional simplices $a_{{\cV}_m}a_{{\cV}_i}$.

In the work \cite{vergeles2023another} it is shown that in $\R^4$ there exist 1-forms
$\omega_{\mu}(x)$ and $\hat{e}_{\mu}(x)$  such that the equalities
\begin{gather}
\omega_{\mu}\left(\frac{1}{2}\,(x_{{\cV}_m}+
x_{{\cV}_i})\,\right)\d
x^{\mu}_{{\cV}_m{\cV}_i}=\omega_{{\cV}_m{\cV}_i},
\label{intr160}
\end{gather}
\begin{gather}
\hat{e}_{\mu}\left(\frac{1}{2}\,(x_{{\cV}_m}+
x_{{\cV}_i})\,\right)\d x^{\mu}_{{\cV}_m{\cV}_i}=
\hat{e}_{{\cV}_m{\cV}_i}.
\label{intr190}
\end{gather}
hold.

Let us write down the long-wavelength limit of the action (\ref{Action_4D}):
\begin{gather}
{\mA'}_g\longrightarrow i\mA_g,  \quad \mA_g=-\frac{1}{4\,l^2_P}\varepsilon_{abcd}\int\mR^{ab}\wedge e^c\wedge e^d,
\nonumber \\
\frac12\sigma_{ab}\mR^{ab}=\frac12\sigma_{ab}\mR^{ab}_{\mu\nu}\d x^{\mu}\wedge\d x^{\nu}
=\big(\partial_{\mu}\omega_{\nu}-\partial_{\nu}\omega_{\mu}+
[\omega_{\mu},\,\omega_{\nu}\,]\big)\d x^{\mu}\wedge\d x^{\nu},
\label{Long_Wav_Grav_Act}
\end{gather}
\begin{gather}
{\mA'}_{\Psi}\longrightarrow i\mA_{\Psi}, \quad \mA_{\Psi}=\frac16\varepsilon_{abcd}\int\Theta^a\wedge e^b\wedge e^c\wedge e^d,
\nonumber \\
\Theta^a=\frac{i}{2}\left[\overline{\Psi}\gamma^a{\cal D}_{\mu}\,\Psi-
\left(\overline{{\cal D}_{\mu}\,\Psi}\right)\gamma^a\Psi\right]\d x^{\mu}, \quad
{\cal D}_{\mu}=\left(\partial_{\mu}+\omega_{\mu}\right),
\label{Long_Wav_Dir_Act}
\end{gather}
\begin{gather}
\mA'_{\Lambda_0}\longrightarrow i\mA_{\Lambda_0}, \quad
\mA_{\Lambda_0}=-\frac{2\Lambda_0}{l_P^2}\int e^0\wedge e^1\wedge e^2\wedge e^3.
\label{Long_Wav_Lambda_Act}
\end{gather}
All other terms in such a transition will contain additional factors to the positive power $(l_P/\lambda)\longrightarrow0$, and therefore they are omitted. Here $\lambda$ is the characteristic wavelength of the physical subsystem. This situation is typical when passing to the long-wave limit in any lattice theory.

The action (\ref{Long_Wav_Grav_Act})-(\ref{Long_Wav_Lambda_Act}) is the Hilbert-Einstein action minimally coupled to the Dirac field and written in Palatini form. It is invariant under diffeomorphisms. This fact is not accidental, since in (\ref{intr110}) the method of introducing coordinates is such that already at this stage the independence of the action from the arbitrariness of introducing coordinates is visible.
We say $"$almost arbitrary$"$, since diffeomorphisms are not arbitrary changes of coordinates, but locally one-to-one and differentiable the required number of times. It is important that the small terms in the long-wave limit, proportional to positive powers of the quantity $(l_P/\lambda)$, are also invariant with respect to diffeomorphisms.

For clarity, we point out that on the lattice all variables and constants are dimensionless and of order one. In particular, the constant $l_P'\sim1$ in (\ref{Latt_Action_Grav}) and (\ref{Latt_Action_Lambda}) is dimensionless, as are the differentials $\d x^{\mu}_{{\cV}_m{\cV}_i}$ in (\ref{intr120}). When passing to dimensional quantities, we assume
\begin{gather}
\d x^{\mu}_{{\cV}_m{\cV}_i}=\d x^{\mu}/l_P\sim1,
\label{Long_Wav_Coord}
\end{gather}
where the differential $\d x^{\mu}$ is measured in centimeters and $l_P\sim10^{-32}\mbox{cm}$. From (\ref{Long_Wav_Coord}) it is seen that the step of the irregular lattice has a size of the order of $l_P$.
And all the terms of the action are dimensionless, but the variables and constants acquire dimensions. For example, in the term (\ref{Long_Wav_Lambda_Act}) the cosmological constant $\Lambda_0\sim l_P^{-2}$.

In the Minkowski signature, the PT-symmetry of the action is determined by the formulas (\ref{PT_transform})-(\ref{PT_trans_Dir_Alg}), (\ref{PT_trans_Ferm_Bil_Fjrm}) with the only difference that in these formulas the replacement should be made $\Psi^{\dag}\longrightarrow\overline{\Psi}$.

\section{High temperature PT-symmetric and low temperature asymmetric phases}

A detailed proof of the fact that in the studied lattice model at ultra-high temperatures a symmetrical phase relative to PT transformation is realized is contained in the work
\cite{vergeles2023another}. There was also an estimate of the temperature at the point of phase transition from the symmetrical to the asymmetrical phase:
\begin{gather}
T_c\sim\frac{\hbar c}{l_P}\sim10^{18} \mbox{GeV} \quad \mbox{or} \quad T_c\sim\left(10^{31}\right)^{\circ}\mbox{K}.
\label{Temperatur_Crit}
\end{gather}
The phase transition temperature can also be estimated as the energy of the Dirac sea enclosed in the Planck volume $V_P\sim l_P^3$: $T_c\sim(\hbar c/l_P^4)l_P^3\sim\hbar c/l_P$. This temperature is of the same order of magnitude as the temperature of the Grand Unification.

Here are just some of the necessary arguments and formulas.

Suppose that a 4D lattice has two 3D sublattices $\Sigma_1$ and $\Sigma_2$ that form its boundary.  For simplicity, we will assume that between $\Sigma_1$ and $\Sigma_2$ there are $N<\infty$  4D lattice layers. If the sublattices $\Sigma_1$ and $\Sigma_2$ are identical (as expected), then, in principle, the partition function can be calculated.

Let the variables $\Psi^{\dag}_{1{\cV}}$ and $\Psi_{2{\cV}}$ be defined on $\Sigma_1$ and $\Sigma_2$, respectively, and $\Phi_{ \xi}\{\ldots\}$ denotes a holomorphic function of fermionic variables on either $\Sigma_1$ or $\Sigma_2$. For simplicity, we assume that all coefficients in these functions are real. Let us accept the following notation: $\Phi_{1\xi}\equiv\Phi_{\xi}\{\Psi^{\dag}_{1{\cV}}\}$,  $\Phi_{2\xi}^{\dag}\equiv\left(\Phi_{\xi}\{\Psi^{\dag}_{2{\cV}}\}
\right)^{\dag}=\Phi_{\xi}\{\Psi_{2{\cV}}\}$. The index $\xi$ lists the independent orthonormal functions from their complete set. To calculate the trace in fermion variables, the functional
\begin{gather}
\Phi_{2\xi}^{\dag}\exp(\beta\mA)\Phi_{1\xi}
\label{product}
\end{gather}
must be placed under the integral (\ref{Partition_function}) and the sum over $\xi$ must be calculated.
On the corresponding 1-simplices of subcomplexes $\Sigma_1$ and $\Sigma_2$, the variables $\{\Omega\}$ are identified in the integral (\ref{Partition_function}). The same rule applies to variables $\{e^a\}$.
In (\ref{product}), the parameter $\beta\equiv1/T\ll1$ is the inverse temperature.

The following statement holds:

{\it Statement.} In a certain finite neighborhood of the point $\beta=0$, the free energy of the partition function (\ref{Partition_function}), with the exception of the term of the form $(\const\cdot{\mN}\cdot\beta^{-1}\ln\beta )$, is a holomorphic function of the variable $\beta$. All action symmetries (\ref{Action_4D}), including discrete PT symmetries, are conserved. $\square$ \\

Let us formulate an important conclusion from the Statement.

Let's place the value $\Phi_{2\xi}^{\dag}{\Theta}^a_{{\cV}_1{\cV}_2}\exp(\beta\mA)\Phi_{1\xi}$
(compare with (\ref{product})) under the integral (\ref{Partition_function}) and calculate only the integral over the Dirac fields and over the variables $\{e\}$.
In the work \cite{vergeles2023another} it was shown that this integral is equal to zero provided that the Statement is true. This result is written as $\langle{\Theta}^a_{{\cV}_1{\cV}_2}\rangle_{\Psi,\,e}=0$, but for further purposes it is more convenient to write
\begin{gather}
\langle{\Theta}^a_{{\cV}_1{\cV}_2}\rangle_{\mbox{Gauge Fix}}=0,  \quad
\left\langle e_{{\cV}_1{\cV}_2}^a\right\rangle_{\mbox{Gauge Fix}}=0.
\label{Mean_Zero_4D}
\end{gather}
The subscript in the equalities (\ref{Mean_Zero_4D}) indicates that the full integral
(\ref{Partition_function}) is calculated with a (locally) fixed gauge. Otherwise, any gauge-non-invariant quantity would automatically vanish under the integral sign. The second equality in (\ref{Mean_Zero_4D}) is obtained in the same way as the first, and it is also a consequence of the fact that under the mean sign we have $\langle e^a_{{\cV}_1{\cV}_2}\rangle\sim\langle{\Theta}^a_{{\cV}_1{\cV}_2}\rangle$
\cite{vladimirov2012phase,volovik2021dimensionless}.
The last equality is in agreement with the fact that the independent quantities $e^a_{{\cV}_1{\cV}_2}$ and ${\Theta}^a_{{\cV}_1{\cV}_2}$ are transformed identically under the action of all symmetries (compare (\ref{Variables_Grav}) and (\ref{Dir_Bil_Form_Trans}), (\ref{Gauge_Trans}) and (\ref{Teta_Gauge_Trans}), (\ref{PT_transform}) and (\ref{PT_trans_Ferm_Bil_Fjrm})).

When the temperature decreases, a phase transition occurs, the essence of which is the birth of space-time.
This means that a non-zero mean appears
$\left\langle e_{{\cV}_1{\cV}_2}^a\right\rangle_{\mbox{Gauge Fix}}\neq0$, the Minkowski signature appears and the inflation phase begins. In this phase, the system is fundamentally described by the action
(\ref{Long_Wav_Grav_Act})-(\ref{Long_Wav_Lambda_Act}). The theory of gravity is non-renormalizable, in which quantum fluctuations grow according to a power law toward short waves. Thus, quantum fluctuations rapidly decrease in the long-wave limit. Therefore, the field of the tetrad $e^a_{\mu}$ can be considered "frozen" or classical. Dirac quantized fields fluctuate against the background of the classical gravitational field.

The conserved operator of the number of fermion particles is defined by the formula
\begin{gather}
{\cal N}=\frac{1}{3!}\varepsilon_{abcd}\int_{\Sigma}(\overline{\Psi}\gamma^a\Psi)e^b\wedge e^c\wedge e^d.
\label{Fermi_Charge_Conserved}
\end{gather}
The operator conservation law (\ref{Fermi_Charge_Conserved}) is a consequence of the equality $\nabla_a\langle \overline{\Psi}\gamma^a\Psi\rangle=0$, which is derived directly from the functional integral by slightly varying the variables $\Psi\longrightarrow e^{i\alpha}\Psi$,
$\overline{\Psi}\longrightarrow e^{-i\alpha}\overline{\Psi}$, $\alpha\longrightarrow0$. Since the same manipulations are possible on the lattice, the conservation of the operator (\ref{Fermi_Charge_Conserved}) is an exact law.

Since $\hat{\cal U}_{PT}^{-1}(\overline{\Psi}\gamma^a\Psi)\hat{\cal U}_{PT}=\overline{\Psi}\gamma^a\Psi$ according to (\ref{PT_transform}), then from (\ref{Fermi_Charge_Conserved}) we have
\begin{gather}
\hat{\cal U}_{PT}^{-1}{\cal N}\hat{\cal U}_{PT}=-{\cal N}.
\label{N_PT}
\end{gather}

Let us point out a fundamental difference between the calculation of the number of fermion particles adopted here and that used in traditional quantum field theory. In the latter case, the number of particles means the number of real particles minus the number of antiparticles, that is, only excitations above the vacuum are taken into account. In our case, only for the state $|0;False\rangle$ satisfying the condition $\Psi|0;False\rangle=0$ (false vacuum in the language of traditional quantum field theory), we have ${\cal N}|0;False\rangle=0$. Such a calculation of the number of fermions is not only possible in lattice theory, but is also necessary in our case. For the vacuum in the traditional field theory $|0\rangle$ we have ${\cal N}|0\rangle=N|0\rangle$, where $N$ is the number of all filled single-particle states of the Dirac sea. In general, the value of the operator ${\cal N}$ will be equal to the number of all filled states of the single-particle Dirac operator.

From Eq. (\ref{N_PT}) it follows that in the high-temperature PT-symmetric phase we have
\begin{gather}
\langle\,{\cal N}\,\rangle=0.
\label{Zero_Ferm_Number}
\end{gather}
Since the operator ${\cal N}$ is conserved, the last equality is valid in all phases.

We assume that in the high-temperature phase the entropy $S$ is maximum with respect to the variable ${\cal N}$:
\begin{gather}
\partial S/\partial{\cal N}=-\beta\mu=0.
\label{Chemical_Potential}
\end{gather}
Here $\mu$ denotes the chemical potential of Fermi particles. Note that the equality (\ref{Chemical_Potential}) does not mean that the entropy is at its absolute maximum, since entropy also depends on other parameters. The entropy (increasing with time) in de Sitter space is presented in \cite{volovik2024sitter}.

Let $|A+\rangle$ be some quantum state of the system defined on a spatially similar hypersurface $\Sigma$.
Let us introduce normal coordinates $x^{\mu}$ with the center at the point $p\in\Sigma$. In a small neighborhood $U$ of this point, the normal coordinates are almost Cartesian, the values of the tetrad and connection are close to the values
\begin{gather}
e^a_{\mu}\approx\delta^a_{\mu}, \quad \omega^{ab}_{\mu}\approx0.
\label{Normal_Coord}
\end{gather}
We will assume that the hyperplane  $x^0=0$ is tangent to the hypersurface $\Sigma$ at the point $p$. Normal coordinates can always be defined in this way. In the neighborhood of point $p$, 4-vector $e^a_{\mu}\d x^{\mu}$ has component $e^0_{\mu}\d x^{\mu}>0$ only if $\d x^0>0$. Note that the property
$e^0_{\mu}\d x^{\mu}>0$ is an invariant under continuous Lorentz transformations (or gauge group) of the 4-vector $e^a_{\mu}\d x^{\mu}$. The triad $e^{\alpha}_i\sim\delta^{\alpha}_i$ also cannot be reduced to the form $(-\delta^{\alpha}_i)$ using continuous elements of the group of three-dimensional rotations.
In the neighborhood of point $p$ the Dirac Hamiltonian has a simple form
${\cal H}_{\Psi}=-i\gamma^0\gamma^{\alpha}\partial_{\alpha}$.

We define the state $|A-\rangle$ as
\begin{gather}
|A-\rangle=\hat{\cal U}_{PT}|A+\rangle.
\label{PT-state}
\end{gather}
Dynamic variables of the v.f. $|A-\rangle$ are defined on elements of the same lattice on which the variables of the v.f. $|A+\rangle$ are defined, but take on different values. Since in both cases the same lattice is used and the coordinates are tied to the lattice vertices, it is also convenient to use the same normal coordinates $x^{\mu}$ in the case of the v.f. $|A-\rangle$. Then instead of (\ref{Normal_Coord}) we have $e^a_{\mu}\approx-\delta^a_{\mu}, \ \omega^{ab}_{\mu}\approx0$ in the neighborhood of point $p$ for the v.f. $|A-\rangle$.

If ${\cal N}|A+\rangle=N_+|A+\rangle$, then according to (\ref{N_PT}) and (\ref{PT-state})
\begin{gather}
{\cal N}|A-\rangle\equiv N_-|A-\rangle=
\hat{\cal U}_{PT}(\hat{\cal U}_{PT}^{-1}{\cal N}\hat{\cal U}_{PT})|A+\rangle
=-\hat{\cal U}_{PT}{\cal N}|A+\rangle=-N_+|A-\rangle.
\nonumber
\end{gather}
In the PT-asymmetric phase, the v.f.
\begin{gather}
|\,\rangle=|A+\rangle+|B-\rangle
\label{PT-nonsymmetric_state}
\end{gather}
is realized with the conditions $\langle A+|A+\rangle=\langle B-|B-\rangle=1$. The constraint (\ref{Zero_Ferm_Number}) leads to the following relation:
\begin{gather}
N_-=-\frac{1+\langle B-|A+\rangle}{1+\langle A+|B-\rangle}N_+.
\nonumber
\end{gather}
If $\langle A+|B-\rangle=\langle B-|A+\rangle$, then $N_+=-N_-$.

For clarity, we need to describe the states  $|A\pm\rangle$ at low temperatures in terms of occupation numbers. To do this, we write out in the neighborhood of $U$ in normal coordinates the equation for the eigenfunctions of the single-particle Dirac Hamiltonian: ${\cal H}_{\Psi}\psi_n=\epsilon_n\psi_n$.
Let us denote the orthonormal wave functions by $\psi_n^{(+)}$ for $\epsilon_n>0$ and $\psi_n^{(-)}$ for $\epsilon_n<0$. The matrix $\gamma^0\gamma^5$ transforms these v.f. into each other, so that between $\psi_n^{(+)}$ and $\psi_n^{(-)}$ there is a one-to-one correspondence. Let us expand the Dirac field operator: $\Psi(x)=\sum_n(a_n\psi_n^{(+)}(x)+b_n\psi_n^{(-)}(x))$, where the operators $\{a_n,\,b_n\}$, as well as their conjugates, are Fermi operators with standard anticommutation properties. The ground state $|0+\rangle$ is determined by the conditions
\begin{gather}
a_n|0+\rangle=0, \quad b^{\dag}_n|0+\rangle=0.
\label{Ground_State+}
\end{gather}

Let us consider the transformation of the operator $\Psi$ under the action of the anti-unitary operator $\hat{\cal U}_{PT}$:
\begin{gather}
\hat{\cal U}_{PT}^{-1}\Psi\hat{\cal U}_{PT}=
\sum_n\bigg[\hat{\cal U}_{PT}^{-1}a_n\hat{\cal U}_{PT}\left(U_{PT}\big(\overline{\psi}_n^{(+)}\big)^t\right)
+\hat{\cal U}_{PT}^{-1}b_n\hat{\cal U}_{PT}\left(U_{PT}\big(\overline{\psi}_n^{(-)}\big)^t\right)\bigg]
\nonumber \\
=\sum_n\bigg[\hat{\cal U}_{PT}^{-1}a_n\hat{\cal U}_{PT}\psi^{(-)}_n
+\hat{\cal U}_{PT}^{-1}b_n\hat{\cal U}_{PT}\psi^{(+)}_n\bigg].
\label{PT_Trans_Psi}
\end{gather}
Here it was taken into account that $U_{PT}\left(\overline{\psi}_n^{(\pm)}\right)^t=\psi_n^{\mp}$. The last equality is verified directly. Since the field (\ref{PT_Trans_Psi}) has the same spatio-temporal transformation properties as the field $\Psi$, the following conclusion should be drawn:
\begin{gather}
\hat{\cal U}_{PT}^{-1}a_n\hat{\cal U}_{PT}=b_n, \quad \hat{\cal U}_{PT}^{-1}b_n\hat{\cal U}_{PT}=a_n,
\nonumber \\
\hat{\cal U}_{PT}^{-1}a^{\dag}_n\hat{\cal U}_{PT}=b_n^{\dag}, \quad \hat{\cal U}_{PT}^{-1}b_n^{\dag}\hat{\cal U}_{PT}=a_n^{\dag}.
\label{PT_Trans_Create_Ann}
\end{gather}
The second line is obtained similarly. Using (\ref{Ground_State+}), (\ref{PT_Trans_Create_Ann}) and the definition $|0-\rangle=\hat{\cal U}_{PT}|0+\rangle$, we find:
\begin{gather}
a_n^{\dag}|0-\rangle=\hat{\cal U}_{PT}b^{\dag}_n|0+\rangle=0,
\nonumber \\
b_n|0-\rangle=\hat{\cal U}_{PT}a_n|0+\rangle=0.
\label{Ground_State-}
\end{gather}
Allowing for a free speech, we can say that the state $|0+\rangle$ is a Dirac vacuum, and the state $|0-\rangle$ is an anti-Dirac vacuum. However, the energies of these states are the same and negative (here the energies of the Dirac seas are not crossed out). In addition, for any local operator $\hat{\cal O}$ or a sum of such (for example, the Hamiltonian) we have
\begin{gather}
\langle0+|\hat{\cal O}|0-\rangle=0,
\label{Matrix_Elem+-}
\end{gather}
because
\begin{gather}
a^{\dag}_na_n|0+\rangle=0, \quad a^{\dag}_na_n|0-\rangle=|0-\rangle,
\nonumber \\
b^{\dag}_nb_n|0+\rangle=|0+\rangle, \quad b^{\dag}_nb_n|0-\rangle=0.
\label{Different_Filing+-}
\end{gather}

\section{Origin of baryon asymmetry}

Let the vacuum in the traditional quantum field theory in Minkowski space be degenerate, i.e. there are several vacua $|v\rangle$ with different field values. Then in the case $v\neq v'$ the following statement holds (compare with (\ref{Matrix_Elem+-})): $\langle v|\hat{\cal O}|v'\rangle=0$, the proof of which is based on the fact that space is infinite \cite{weinberg1995quantum,volovik2007quantum,sinai1983theory}. In our theory, equality (\ref{Matrix_Elem+-}) is also possible only on an infinite lattice, since only on infinite lattices is a phase transition and a qualitative separation of states into states of the form $|\pm\rangle$ possible. Thus, one should study physics over one of the vacua $|v\rangle$, but not over their superposition.

The general picture studied here is qualitatively more complex due to the dynamism. As a result of the phase transition from the PT-symmetric to the asymmetric phase, the wave-function of the system takes the form (\ref{PT-nonsymmetric_state}). Indeed, in the integral (\ref{Partition_function}) the integration over $e^a_{{\cV}_1{\cV}_2}$ is within the limits $(|e_{{\cV}_1{\cV}_2}|<1)$, and the Dirac variables are interchangeable according to (\ref{PT_transform}). Since a non-zero order parameter $e^a_{\mu}$ appears ($e^a_{\mu}\sim\delta^a_{\mu}$ in the state $|A+\rangle$ and
$e^a_{\mu}\sim-\delta^a_{\mu}$ in the state $|B-\rangle$), then time $\d t=e^0_{\mu}\d x^{\mu}$ appears. In these states, time flows in opposite directions. During the initial quantum of time $t_P$, interactions between the specified states take place, so that the number of fermions $N_+=-N_-$ can change. But after time $t_P$, any interaction between states $|A+\rangle$ and $|B-\rangle$ ceases, and the accumulated value $N_+$ is preserved (compare with the example from the Introduction). This value of $N_+$ is preserved, not being, generally speaking, the equilibrium value of $N_+^{(0)}$ at any temperature and zero value of the chemical potential $\mu$. Note that in the state $|0+\rangle$ (\ref{Ground_State+}) $N_+^{(0)}$ is realized at zero temperature.

Let us give the simplest illustration of the definition of the number $N_+^{(0)}$ using the example of an ideal Fermi gas consisting of two degrees of freedom with energies $\varepsilon=\pm\epsilon$ and zero chemical potential. The equilibrium number of particles at the $\varepsilon$ level is $n(\varepsilon)=(e^{\beta\varepsilon}+1)^{-1}$.
The total number of particles $N^{(0)}=n(-\epsilon)+n(\epsilon)=1$ at any temperature. In other words, the number of excitations (electrons) $n(\epsilon)$ is equal to the number of holes (positrons) $1-n(-\epsilon)$. In the case of $\beta\rightarrow0$ we have $n(-\epsilon)\rightarrow(1/2+0)$, $n(\epsilon)\rightarrow(1/2-0)$, or $<\varepsilon>\rightarrow0$. $N^{(0)}$ is analogous to $N_+^{(0)}$.

Our goal is to estimate, at least qualitatively, the value $\delta N_+\equiv N_+-N_+^{(0)}$. For a fluctuation to be considered thermodynamic, the time interval $\tau$ must be \cite{landau1980lifshitz}
\begin{gather}
\tau\gg\frac{\hbar}{T}.
\label{Thermodynamic_Time}
\end{gather}
Here the temperature is close to the phase transition temperature $T_c\sim\hbar c/l_P$ (\ref{Temperatur_Crit}). Then using (\ref{Thermodynamic_Time}) we obtain the estimate: $\tau\gg l_P/c\sim t_P$. Further arguments are given in favor of the fact that the equality (\ref{Matrix_Elem+-}) already holds for $\tau\geq t_P$, although this assumption is rather a hypothesis. Therefore, the mechanism of thermodynamic fluctuations does not work here. The problem of thermodynamic fluctuations in the vacuum of de Sitter space was studied in the review \cite{volovik2024thermodynamics}, where interesting results were obtained.

We assume that a first-order phase transition occurs here. The basis for this assumption is the following fact: in the asymmetric phase $N_+=-N_-\neq0$, and the fluctuations of the number $N_+$ end during the first quantum of time $t_P$ (see below). This means that a non-zero number $N_+$ arises almost instantly as a result of the phase transition. During the first-order phase transition, local embryos of the form $|\Xi\rangle=|\Xi+\rangle+|\Xi-\rangle$ are formed (compare with (\ref{PT-nonsymmetric_state})). By assumption, the w.f. $|\Xi\rangle$ depends on variables defined on several adjacent lattice elements. The number of such variables is of the order of $10^n, \,n\sim1$. Under this assumption, the complete w.f. of the system is approximately represented as
\begin{gather}
|\,\rangle\approx\prod_{\Xi}|\Xi\rangle.
\label{Quantum_Fluct_1}
\end{gather}
To estimate the value of $\delta N_+$ as a result of quantum fluctuations, we will use the formulas
\begin{gather}
|\delta E_{\Xi+}\delta N_{\Xi+}|\sim\hbar|\dot{N}_{\Xi+}|, \quad
|\delta E_{\Xi+}|t_p\sim\hbar,
\label{Quantum_Fluct_2}
\end{gather}
\begin{gather}
\delta N_+\approx\sum_{\Xi}\delta N_{\Xi+} \longrightarrow
|\delta N_+|\sim\sqrt{N_+}|\delta N_{\Xi+}|.
\label{Total_Result}
\end{gather}
The last relation in (\ref{Total_Result}) follows from the fact that the sign of each fluctuation $\delta N_{\Xi+}$ is random and the order of magnitude of all these fluctuations is the same. Here $E_{\Xi+}$ and $N_{\Xi+}$ denote the fermion contribution to the energy and the number of particles of the state $|\Xi\rangle$. Further, $|\dot{N}_{\Xi+}|\sim\nu|N_{\Xi+}|/t_P$, $\nu\ll1$. The smallness of the parameter $\nu$ means that during the first quantum of time $t_P$ the value of the number $N_{\Xi+}$ changes insignificantly. Combining the last estimates and relations (\ref{Quantum_Fluct_2}), (\ref{Total_Result}), we find:
\begin{gather}
|\delta N_+|\sim\nu\sqrt{N_+}N_{\Xi+}\sim\nu\sqrt{N_+},
\label{Ansver}
\end{gather}
since in this problem we can set $N_{\Xi+}\sim1$.

Here we can see the fundamental difference in the role of the type of phase transition in the traditional theory of baryosynthesis \cite{rubakov1996electroweak,kuzmin1985anomalous} and in our approach. In the first case, the first type guarantees the appearance of embryo and bubbles with broken symmetry, the moving domain walls of which stimulate baryosynthesis. In our theory, the first-order transition leads to the factorization formula (\ref{Quantum_Fluct_1}), and then to subsequent formulas (\ref{Quantum_Fluct_2})-(\ref{Ansver}).

To estimate the value of $\nu$, it is necessary to estimate the amplitude of the transition of one fermion from the state $|\Xi+\rangle$ to the state $|\Xi-\rangle$ during a minimal quantum of time. In lattice theory, this implies calculating the value
\begin{gather}
\nu\sim\left|\langle\Xi-|\mA^{(1)}_{\Psi}|\Xi+\rangle\right|^2,
\label{Estim_nu_2}
\end{gather}
where $\mA^{(1)}_{\Psi}$ denotes the part of the fermionic action (\ref{Latt_Action_Ferm}) defined on a space-like subcomplex of thickness one layer in the time direction. On the right side in (\ref{Estim_nu_2}) the part of the amplitude that increases (decreases) the number of particles in the state $|\Xi+\rangle$ is highlighted. Note that due to the high temperature, the states $|\Xi+\rangle$ and $|\Xi-\rangle$ are a superposition of states with different numbers $N_{\Xi+}$ and $N_{\Xi-}$, but with the condition $N_++N_-=0$.

There are three sources of smallness of the quantity (\ref{Estim_nu_2}): 1) smallness of the quantity ${\Theta}^a_{{\cV}_1{\cV}_2}$ near the phase transition point (see
(\ref{Mean_Zero_4D})); 2) the beginning of the process of building up the Dirac vacuum in the state
$|\Xi+\rangle$ and the "anti-Dirac vacuum" in the state $N_{\Xi-}$. At the end of this process we have strict equality (\ref{Matrix_Elem+-}), but already at the very beginning this factor plays a decisive role.
3) Another source of smallness of the parameter $\nu$ is the overlap integral when integrating over the variables $e^a$. Unfortunately, we are unable to give a reliable estimate of the value $\nu$, limiting ourselves to merely stating the fact $\nu\ll1$.

Let us give a numerical estimate and compare it with observations. Let us estimate the number $N_+$ based on the current state of the visible part of the Universe: $N_+\sim(L_0/l_P)^3\sim10^{181}$. Here $L_0\sim2\cdot10^{28}\mbox{cm}$ is the size of the visible part of the Universe and $l_P\sim10^{-32}\mbox{cm}$ is the Planck scale. Then according to (\ref{Estim_nu_2}) for the density of fermion asymmetry
\begin{gather}
\delta n_+\equiv\nu\sqrt{N_+}/L_0^3\sim\nu(L_0l_P)^{-3/2}\sim\nu\cdot10^6\mbox{cm}^{-3}.
\label{Dens}
\end{gather}
The experimental value is $\delta n_{+(\mbox{exp})}\sim10^{-5}\mbox{cm}^{-3}$. From this it is clear that if $\nu\sim10^{-10}$, then the described mechanism of fermion symmetry breaking works.

\section{Conclusion}

Here we presented only the idea of the emergence of particle-antiparticle asymmetry at the earliest stage of the existence of the Universe, preceding the inflation phase. The described scenario is realized in the discrete (lattice) theory of gravity.

It is necessary to compare Sakharov's conditions for baryogenesis with those conditions that take place in the model under study. Let us write out Sakharov's conditions: 1) violation of baryon charge conservation; 2) violation of C- and CP-invariance; 3) absence of thermal equilibrium at the stage of processes with non-conservation of baryon charge. The third condition is automatically satisfied here. Furthermore, in the model under consideration, baryons and leptons are not distinguished. Therefore, the violation of fermion charge is considered here. The total fermion charge is conserved, but redistributed between the states $|+\rangle$ and $|-\rangle$. Each of these states individually is not C-even, and their interaction ceases within a time of the order of $t_P$. In this interpretation, the first two Sakharov conditions are contained here.

According to the author, the obtained result may stimulate the search for a more perfect version of the discrete theory of gravity, better suited for some calculations.

\begin{acknowledgments}

This work was carried out as a part of the State Program 0033-2019-0005.

\end{acknowledgments}


\end{document}